\documentclass[journal=jpccck, manuscript=article]{achemso}

\usepackage[version=3]{mhchem} 
\usepackage{xcolor}
\usepackage{booktabs}
\usepackage{natbib}

\author{Mikhail E. Zaytsev}
\affiliation{Physics of Fluids, Max Planck Center Twente for Complex Fluid Dynamics and J.M. Burgers Centre for Fluid Mechanics, MESA+ Institute for Nanotechnology, University of Twente, P.O. Box 217, 7500AE Enschede, The Netherlands}
\alsoaffiliation{Physics of Interfaces and Nanomaterials, MESA+ Institute for Nanotechnology, University of Twente, 7500AE Enschede, The Netherlands}

\author{Guillaume Lajoinie}
\affiliation{Physics of Fluids, Max Planck Center Twente for Complex Fluid Dynamics and J.M. Burgers Centre for Fluid Mechanics, MESA+ Institute for Nanotechnology, University of Twente, P.O. Box 217, 7500AE Enschede, The Netherlands}
\alsoaffiliation{TechMed Centre, University of Twente, Enschede, The Netherlands}

\author{Yuliang Wang}
\affiliation{Robotics Institute, School of Mechanical Engineering and Automation, Beihang University, Beijing 100191, P.R. China}

\author{Detlef Lohse}
\affiliation{Physics of Fluids, Max Planck Center Twente for Complex Fluid Dynamics and J.M. Burgers Centre for Fluid Mechanics, MESA+ Institute for Nanotechnology, University of Twente, P.O. Box 217, 7500AE Enschede, The Netherlands}
\alsoaffiliation{Max Planck Institute for Dynamics and Self-Organization, Am Fassberg 17, 37077 G\"ottingen, Germany}

\author{Harold J. W. Zandvliet}
\affiliation{Physics of Interfaces and Nanomaterials, MESA+ Institute for Nanotechnology, University of Twente, 7500AE Enschede, The Netherlands}
\email {h.j.w.zandvliet@utwente.nl}

\author{Xuehua Zhang}
\affiliation{Department of Chemical and Materials Engineering, University of Alberta, 12-211 Donadeo Innovation Centre for Engineering, Edmonton, Alberta, Canada}
\alsoaffiliation{Physics of Fluids, Max Planck Center Twente for Complex Fluid Dynamics and J.M. Burgers Centre for Fluid Mechanics, MESA+ Institute for Nanotechnology, University of Twente, P.O. Box 217, 7500AE Enschede, The Netherlands}
\email{xuehua.zhang@ualberta.ca}

\title{Plasmonic Bubbles in n-Alkanes}

\begin{document}



\begin{abstract}

In this paper we study the formation of microbubbles upon the irradiation of an array of plasmonic Au nanoparticles with a laser in n-alkanes ($C_nH_{2n+2}$, with n = 5-10). Two different phases in the evolution of the bubbles can be distinguished. In the first phase, which occurs after a delay time $\tau_{d}$ of about 100 $\mu s$, an explosive microbubble, reaching a diameter in the range from 10 $\mu m$ to 100 $\mu m$, is formed. The exact size of this explosive microbubble barely depends on the carbon chain length of the alkane, but only on the laser power $P_l$. With increasing laser power, the delay time prior to bubble nucleation as well as the size of the microbubble both decrease. In the second phase, which sets in right after the collapse of the explosive microbubble, a new bubble forms and starts growing due to the vaporization of the surrounding liquid, which is highly gas rich. The final bubble size in this second phase strongly depends on the alkane chain length, namely it increases with decreasing number of carbon atoms.
Our results have important implications for using plasmonic heating to control chemical reactions in organic solvents.

\end{abstract}

\section{Introduction}
Microbubbles generated around plasmonic gold nanoparticles by resonant laser irradiation are relevant to various applications in biomedical diagnosis and cancer therapy \cite{lapotko2009, emelianov2009, baffou2013, shao2015, liu2014, fan2014}, solar energy harvesting \cite{neumann2013, neumann2013-1, fang2013, boriskina2013, ghasemi2014, baral2014, guo2017}, micromanipulation of nano-objects\cite{krishnan2009, zhang2011, zhao2014, tantussi2018, xie2017} and locally enhanced chemical reactions \cite{baffou2014,adleman2009}. Controlled and highly tunable plasmonic bubble formation is in particular suitable for colloidal nanoparticles deposition and clusters formation on a substrate to enhance the signal for Raman spectroscopy in ultrasensitive analysis and sensing \cite{lin2016}. Moreover, the plasmonic effect may locally facilitate and enhance catalytic conversion while the entire system still remains at ambient temperature, which allows to utilize the conversion process in miniaturized systems suitable for lab-on-a-chip applications \cite{adleman2009}. 

Owing to the technological relevance of these microbubbles, the dynamics of their growth and shrinkage has been scrutinized in several investigations \cite{neumann2013, baffou2014jpc, carlson2012, fang2013, baral2014, liu2015, wang2017,wang2018}.  Plasmonic bubble generation is a complex phenomenon, which involves numerous physical processes, such as heat transfer from the nanoparticle to the liquid, supersaturation of the liquid, phase transitions, bubble nucleation, diffusion of gas dissolved in the liquid, evaporation of the liquid and many others. The irradiation of metal nanoparticles immersed in liquid with a continuous laser at resonant wavelength results in plasmonic excitation of the nanoparticles and thereby, in a rapid increase of the temperature of the nanoparticles. Consequently, the liquid surrounding the nanoparticles also exhibits an increase in the temperature, albeit much slower than the nanoparticles. In the absence of any nuclei, the surrounding liquid can become superheated, i.e., reach a temperature that exceeds the boiling temperature. Eventually, however, nucleation will occur, resulting in the explosive growth of a vapor microbubble.  In a recent study we have found that the dynamics of plasmonic microbubble nucleation and evolution \textit{in water} can be divided into 4 subsequent phases, namely, an initial explosive vapor bubble phase (life phase 1), an oscillating bubble phase (life phase 2),  a vaporization dominated growth phase (life phase 3), and a gas diffusion dominated growth phase (life phase 4) \cite{wang2017, wang2018}. 

Up to now, the studies have focused on the formation of plasmonic bubbles in pure water and aqueous solutions or suspensions. However, many plamonic assisted processes take place in presence of an organic phase. In order to better understand the physics of the formation and growth of these microbubbles in different liquid media, it is essential to vary the relevant physical properties of the liquid host, such as the boiling point, the enthalpy of vaporization, the heat capacity, the surface tension, the thermal conductivity, and the gas solubility of the host liquid. In this study, our liquids of choice are n-alkanes. These are simple hydrocarbons, which consist of a carbon backbone and their general formula is $C_nH_{2n+2}$. In Table \ref{properties} various physical properties of the n-alkanes and water are summarized. Their properties depend on the carbon chain length, which make them ideally suited for investigating bubble formation in an organic phase. Here we will systematically study the dynamics of plasmonic microbubbles formed on the substrate with an array of gold nanoparticles (GNP) with a diameter of about 100 nm immersed in n-alkane under a continuous laser irradiation ($\lambda$=532 nm).

\begin{table}\centering
\caption{Physical properties of water and n-alkanes at 20$^\circ$ C \cite{nist}.}
\begin{tabular}{lccccccc}
 & $H_2O$ & $C_5H_{12}$ & $C_6H_{14}$ & $C_7H_{16}$ & $C_8H_{18}$ & $C_9H_{20}$ & $C_{10}H_{22}$ \\
\midrule
Boiling point ($K$) & 373.2 & 309.2 & 341.9 & 371.5 & 398.8 & 423.9 & 447.2 \\
Molar mass ($g/mol$) & 18.01 & 72.15 & 86.18 & 100.21 & 114.23 & 128.76 & 142.29 \\
Density ($kg/m^3$) & 998 & 626 & 659 & 684 & 702 & 718 & 730 \\
Heat capacity ($kJ/kg K$) & 4.184 & 2.293 & 2.231 & 2.222 & 2.209 & 2.189 & 2.174 \\
Enthalpy of vap. ($kJ/kg$) & 2193.2 & 360.2 & 337.8 & 316.6 & 304.4 & 288.5 & 281.3 \\
Surface tension ($mN/m$) & 72.74 & 16.01 & 18.49 & 20.55 & 21.64 & 22.90 & 23.82 \\
Viscosity ($mPa s$) & 1.002 & 0.227& 0.312 & 0.411 & 0.541 & 0.697 & 0.912 \\
Thermal cond. ($W/m K$) & 0.598 & 0.113 & 0.128 & 0.132 & 0.126 & 0.128 & 0.131 \\
$N_2$ solubility ($mol/mol$) & 0.017 & 0.559 & 0.449 & 0.374 & 0.324 & 0.281 & 0.252 \\
\bottomrule
\end{tabular}
\label{properties}
\end{table}

We found that, as compared to water as host medium, the dynamics of plasmonic microbubbles in n-alkanes differs significantly. The bubbles in n-alkanes exhibit only two consecutive phases during their evolution, rather than four, as in the water case. In the first phase, which occurs after a delay time $\tau_{d}$ after turning on the laser, still an explosive vapor microbubble is formed. In the second phase, after the collapse of the initial microbubble, a new bubble forms and starts growing due to the vaporization of the surrounding liquid. The life phases 2 and 4 found in the water case are however absent. These results bear important implications for using plasmonic heating to control chemical reactions in organic solvents.

\section{Methods}
\subsection*{Sample Preparation}
A gold layer of approximately 45 nm was deposited on an amorphous fused-silica wafer by using an ion-beam sputtering system (home-built T$^\prime$COathy machine, MESA+ NanoLab, Twente University). A bottom anti-reflection coating (BARC) layer ($\sim$186 nm) and a photoresist (PR) layer ($\sim$200 nm) were subsequently coated on the wafer. Periodic nanocolumns with diameters of approximately 110 nm were patterned in the PR layer using displacement Talbot lithography (PhableR 100C, EULITHA) \cite{the2017}. These periodic PR nanocolumns were subsequently  transferred at wafer level to the underlying BARC layer, forming 110 nm BARC nanocolumns by using nitrogen plasma etching (home-built TEtske machine, NanoLab) at 10 mTorr and 25 W for 8 min. Using these BARC nanocolumns as a mask, the Au layer was subsequently etched by ion beam etching (Oxford i300, Oxford Instruments, United Kingdom) with 5 sccm Ar and 50-55 mA at an inclined angle of $5^{\circ}$. The etching for 9 min resulted in periodic Au nanodots supported on cone-shaped fused-silica features. The remaining BARC was stripped using oxygen plasma for 10 min (TePla 300E, PVA TePla AG, Germany). The fabricated array of Au nanodots was heated to $1100^{\circ}$C in 90 min and subsequently cooled passively to room temperature. During the annealing process, these Au nanodots re-formed into spherical-shaped Au nanoparticles.

\begin{figure*}[h]
 	\includegraphics[width=1\textwidth]{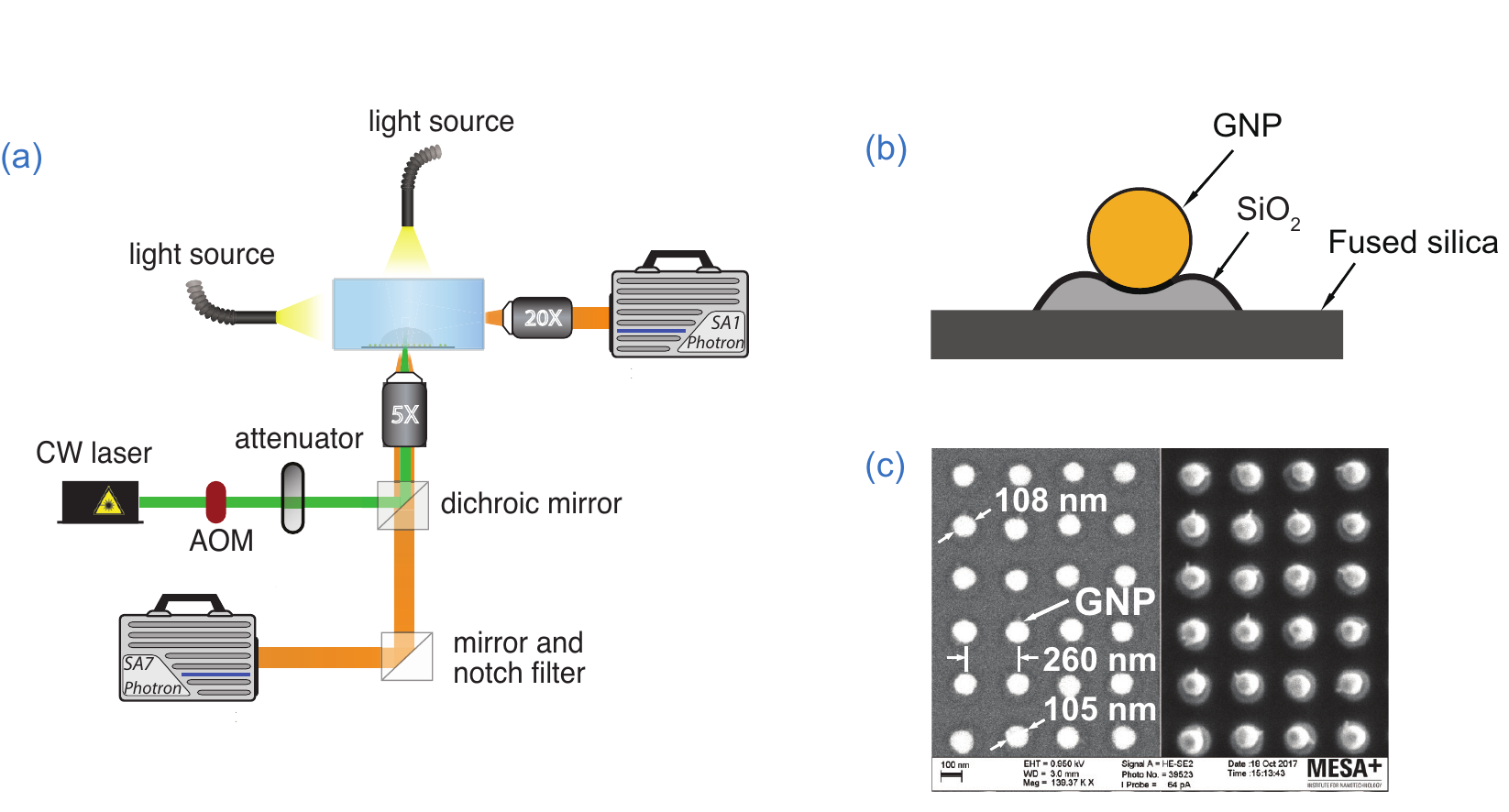}
 	\caption{
   a) Schematic of the optical imaging facilities for plasmonic microbubble formation observation, b) Schematic of a gold nanoparticle sitting on a $SiO_2$ island on a fused-silica substrate, c) SEM images of the patterned gold nanoparticle sample surface. 
 	}
 	\label{fig:experiment}
 \end{figure*}

\subsection*{Setup description}
The experimental setup for plasmonic microbubbles imaging is shown in Figure \ref{fig:experiment}. The gold nanoparticle decorated sample was placed in a quartz glass cuvette and filled with liquid. A continuous-wave laser (Cobolt Samba) of 532 nm wavelength and a maximum power of 300 mW was used for sample irradiation. An acousto-optic modulator (Opto-Electronic, AOTFncVIS) was used as a shutter to control the laser irradiation on the sample surface. 400 $\mu s$ and 4s laser pulses were generated and controlled by pulse/delay generator (BNC model 565) in order to study the short-term and the long-term dynamics of microbubbles, respectively. The laser power was controlled by using a half-wave plate and a polarizer and measured by a photodiode power sensor (S130C, ThorLabs). Two high-speed cameras were installed in the setup, one (Photron SA7) equipped with 5x long working distance objective (LMPLFLN, Olympus) and the other (Photron SA1) equipped with various long working distance objectives: 5x (LMPLFLN, Olympus), 10x (LMPLFLN, Olympus) and 20x (SLMPLN, Olympus) and operated at various framerates from 5 kfps up to 500 kfps. The first camera was used for a top-view and the second one for side-view. Two light sources,  an Olympus ILP-1 and a Schott ACE I provided illumination for the two high-speed cameras. In our experiments n-alkanes from Sigma-Aldrich and MiliQ water were used.

\section{Results and discussion}

 \begin{figure*}[h]
 	\includegraphics[width=1\textwidth]{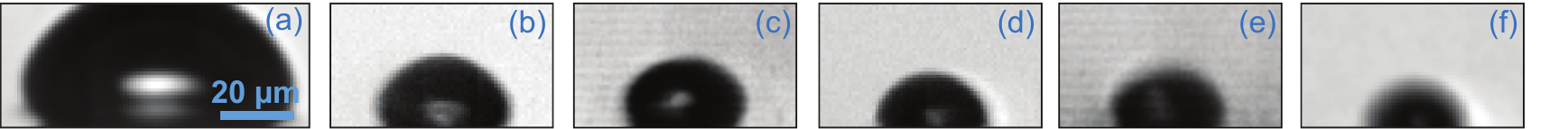}
 	\caption{
 		Examples of initial explosive bubble at maximum expansion in a) water and n-alkanes: b) pentane, c) hexane, d) heptane, e) octane and f) nonane at the same laser power $P_l$ = 60 mW.
 	}
 	\label{fig:1st bubble example}
 \end{figure*}

The initial stages of the explosive growth of plasmonic microbubbles in n-alkanes has been monitored with a high-speed camera operated at a frame rate of up to 500 kfps. Figure \ref{fig:1st bubble example} depicts examples of these vapor microbubbles at their maximum expansion for water, pentane, hexane, heptane, octane and nonane, respectively. Decane is not considered here, as the life cycle of an explosive bubble in decane is too short to be characterized with available frame rates. Figures \ref{fig:r, t, V}(a) and \ref{fig:r, t, V}(b) show the bubble radius $R_b$ and the delay time before nucleation of the bubble $\tau_d$, respectively. With increasing laser power $P_l$, the bubble radius as well as the delay time decrease, similarly to what has been observed for water \cite{wang2018}. The delay time increases with increasing alkane chain length, whereas the maximum microbubble size mainly depends on the laser power.
For a given laser power of 60 mW, the microbubbles formed in water have a radius of about 50 $\mu m$ and are by far the largest, followed by pentane, hexane, heptane, octane and finally nonane. It is worth noting that only a slight difference in bubble size of the various n-alkanes was observed. On first sight this seems counterintuitive, considering the significant variation in boiling temperatures of the alkanes, from 36 $^\circ$C for $C_5H_{12}$ to 151 $^\circ$C for $C_9H_{20}$, but we will explain this later.

 \begin{figure*}[h]
 	\includegraphics[width=1\textwidth]{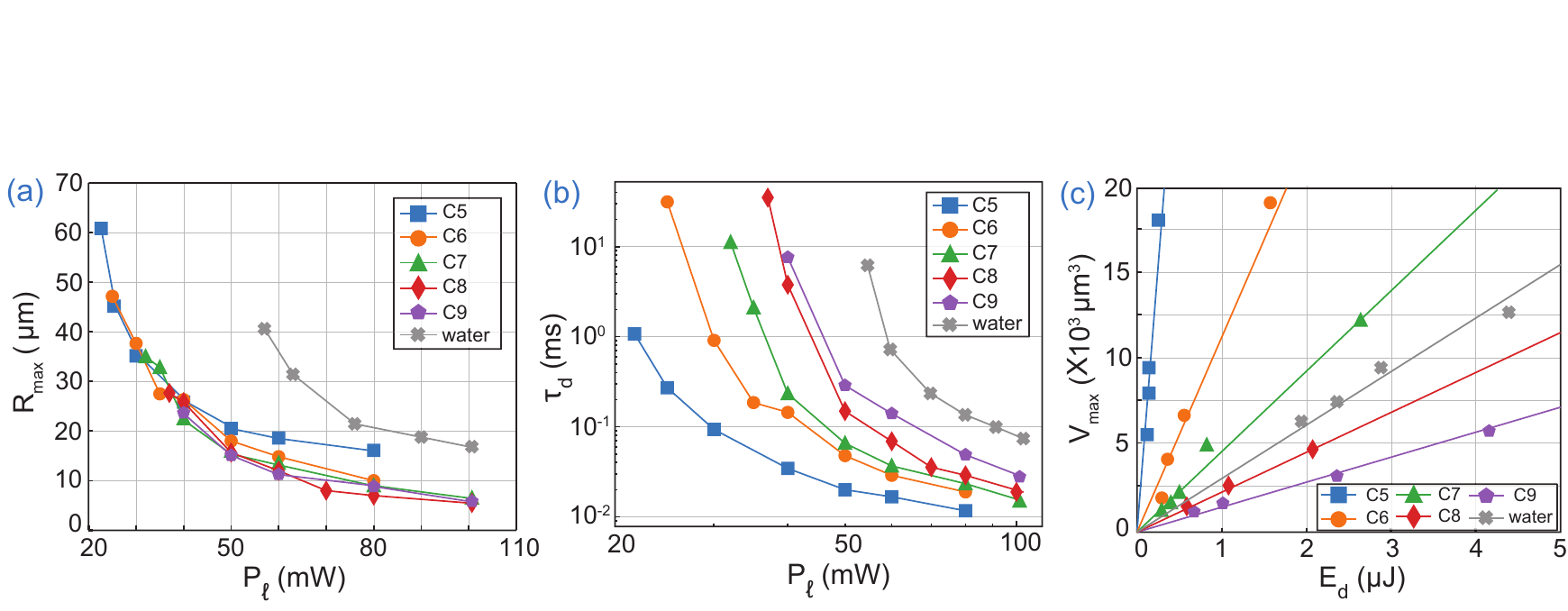}
 	\caption{
 		a) Maximum bubble radius $R_{max}$ versus laser power $P_\ell$. b) Delay time $\tau_d$ prior to bubble nucleation versus laser power $P_\ell$. c) Maximum bubble volume $V_{max}$ versus the deposited energy $E_d=P_\ell \tau_d$. C5, C6, C7, C8 and C9 refer to pentane, hexane, heptane, octane and nonane, respectively. All error bars lie within the size of the markers.
 	}
 	\label{fig:r, t, V}
 \end{figure*}

Figure \ref{fig:r, t, V}(c) shows the volume of the microbubble versus the cumulative energy before nucleation, hereafter referred to as the deposited energy, $E_d=P_\ell \tau_d$. For all liquids the volume of the microbubble scales linearly with the deposited energy. The curve for water in Fig. \ref{fig:r, t, V}(c) is located in between the curves of heptane and octane, which is consistent with the boiling temperatures of these liquids, i.e. 98 $^\circ$C for heptane and 126 $^\circ$C for octane, respectively.

\begin{figure*}[h]
 	\includegraphics[width=1\textwidth]{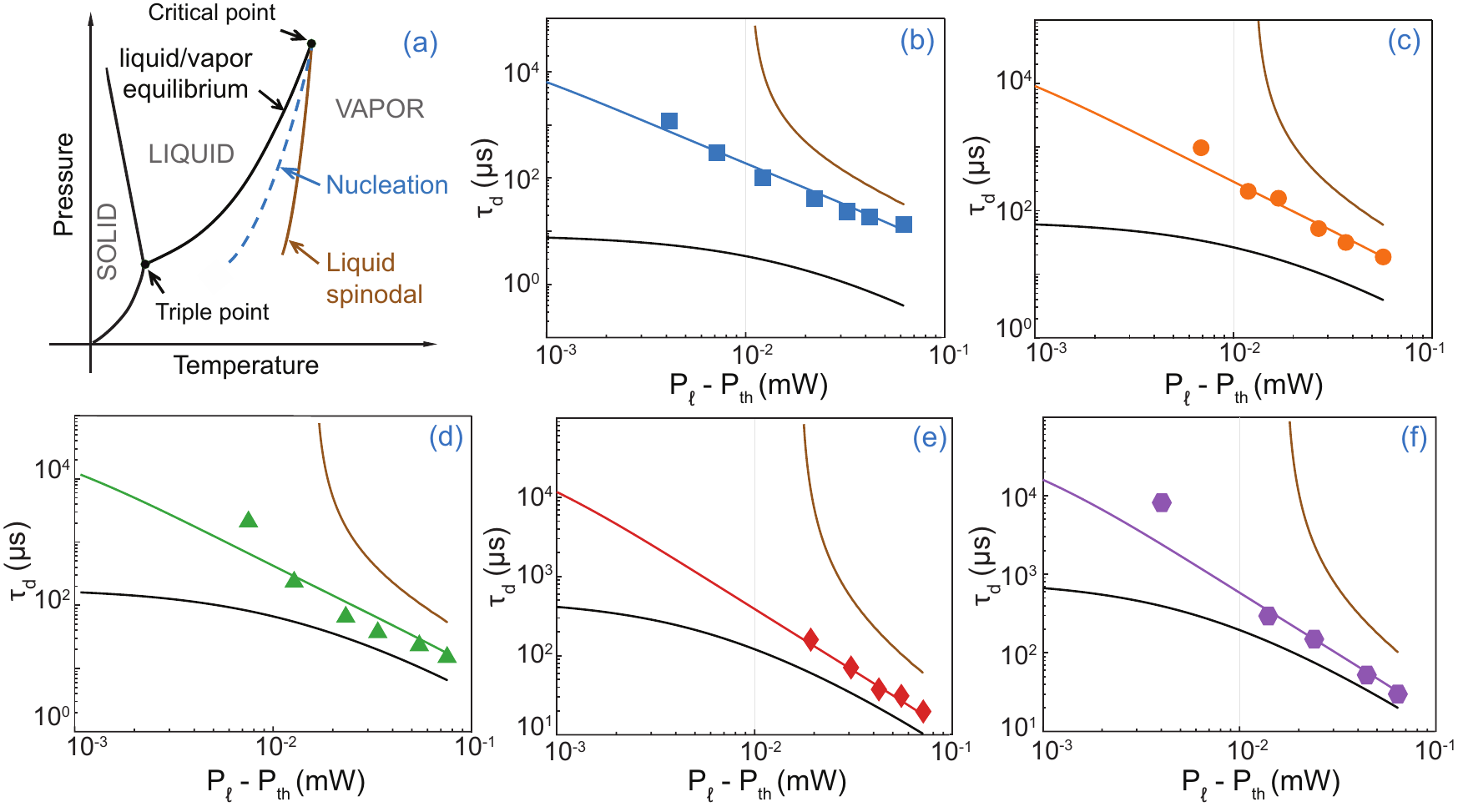}
 	\caption{
 		a) Schematic phase diagram. (b-f) Double logarithmic plot of delay time $\tau_d$ as function of laser power $P_\ell$ for (b) pentane, (c) hexane, (d) heptane, (e) octane and (f) nonane. $P_{th}$ is the threshold laser power below which no nucleation can occur. The symbols represent the experimental data and the solid lines are the spinodal lines (bronze), the equilibrium lines (black), and the fitted curves using Eq. \ref{heat_conduction}. All fitted lines are within the metastable region, i.e. between the equilibrium and spinodal boundaries.
 	}
 	\label{fig:phase diag}
 \end{figure*}

Prior to bubble nucleation, the liquid has to be heated up to a temperature that exceeds its boiling temperature. Particularly, the nucleation temperature of very pure liquids can be much higher than their boiling temperature \cite{skripov1974,skripov1970}. There is, however, a limit to this superheating. For a given pressure, the maximum attainable temperature is the so-called spinodal temperature. At the spinodal temperature the evaporation enthalpy vanishes, resulting in spontaneous homogeneous nucleation. However, in most cases nucleation occurs at a temperature substantially lower than the spinodal temperature, owing to the presence of tiny cracks, cavities or pits filled with gas, impurities or aggregations of gas molecules that can act as centers of nucleation \cite{skripov1974, blander1975, puchinskis2001}. The latter nucleation process is referred to as heterogeneous nucleation. A schematic phase diagram is shown in figure \ref{fig:phase diag} (a).

\begin{figure*}[h]
 	\includegraphics[width=0.5\textwidth]{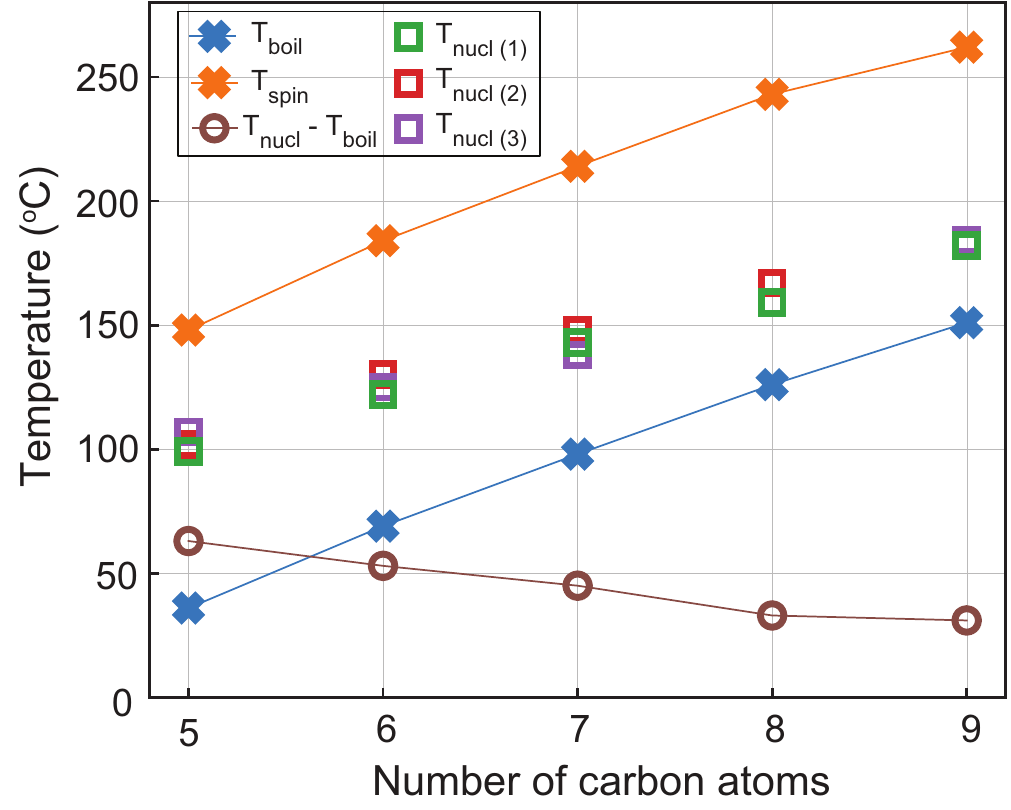}
 	\caption{
 		Boiling, nucleation, spinodal and superheating (difference between average nucleation and boiling temperatures) temperatures for pentane, hexane, heptane, octane, and nonane. The nucleation temperatures form repeating experiments have been plotted to show the reproducibility in the different runs. All error bars lie within the size of the markers.
 	}
 	\label{fig:superheat}
 \end{figure*}

In order to determine the nucleation temperature, just as done in our previous study for water \cite{wang2018}, we numerically calculate the temperature evolution around a nanoparticle, assuming spherical geometry, by solving the heat conduction equation,

\begin{equation}
	\partial_t(T(r,t)) = \frac{p_l(r,t)}{\rho c_p} + \kappa\frac{1}{r^2}\partial_r(r^2\partial_rT(r,t)) ,
	\label{heat_conduction}
\end{equation}

\noindent where $\kappa$, $\rho$, and $c_p$ are thermal diffusivity, density, and heat capacity of water, $r$ is the distance to the GNP, and $p_l(r,t)$ is the deposited power density (unit in $W/m^3$), which is assumed to be constant for a radius $r$ within the GNP, and 0 elsewhere.
The temperature field generated by the nanoparticles array has been calculated by superposition, within a Gaussian beam profile. The resulting time-dependent temperature field is given by

\begin{equation}
	T(x,y,z,t) = \sum_{i=1}^{N}\left(T_i(d_{i,(x,y,z)},t)\right) ,
\end{equation}

\noindent where $T_i$ the temperature field created by the nanoparticle $i$, $d_i$ the distance from the center of the $ith$ nanoparticle. Using the above equations, one directly obtains the time required to reach a given temperature for a given laser power. This approach was used for every alkane to fit the experimental data using a root-mean-square-minimization method, resulting in the solid curves in Fig. \ref{fig:phase diag} (b-f). The nucleation temperatures for the alkanes are 79$^\circ$C, 102$^\circ$C, 123$^\circ$C, 139$^\circ$C and 160$^\circ$C for $C_5H_{12}$, $C_6H_{14}$, $C_7H_{16}$, $C_8H_{18}$ and $C_9H_{20}$, respectively.
In Figure \ref{fig:superheat}, the boiling, nucleation, and spinodal temperatures as well as the superheating temperature (nucleation temperature -- boiling temperature) are shown. Boiling, nucleation and spinodal temperatures all increase with the length of the alkane chain. Notably, the degree of superheat decreases with increasing number of carbon atoms in alkanes from 63 K for pentane to 29 K for nonane.

The bubble nucleation mechanism in alkanes appears to differ from the one in water. Air solubility in water decreases with increasing temperature and, therefore, the heated liquid becomes supersaturated and, therefore, expelled gas molecules facilitate bubble nucleation. However, in n-alkanes the solubility of air \textit{increases} with increasing temperature \cite{battino1984,mizerovsky2010}, causing undersaturation in the heated zone. However, it is important to point out (see Table \ref{properties}) that the amount of dissolved gas in alkanes is 15-30 times more than in water, so the gas level in the surrounding liquid may still play a role in bubble nucleation.

From the above mentioned numerical calculations the threshold values of laser power $P_{th}$ for all alkanes have been extracted, namely $P_\ell ^ {th}$ = 18 mW, 24 mW, 28 mW, 34 mW, 37 mW  for pentane, hexane, heptane, octane and nonane, respectively. The laser powers are consistent with the experimental results (see Figure \ref{fig:r, t, V}), which confirms the robustness of our method. Below these threshold values, heat diffusion prevents the system from reaching the required nucleation temperature and, therefore, a bubble cannot be generated.

\begin{figure*}[h]
 	\includegraphics[width=0.5\textwidth]{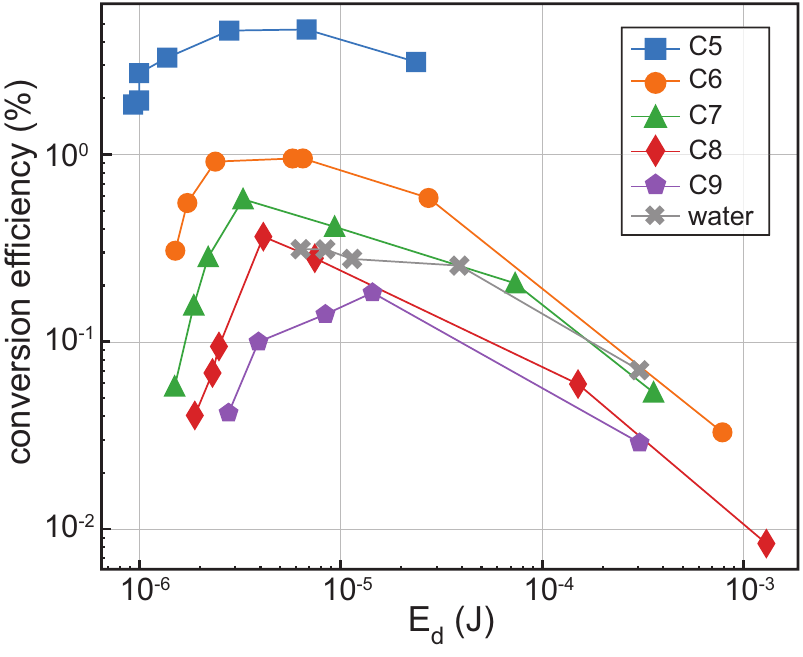}
 	\caption{
 		Conversion efficiency of the GNPs for the energy converted from laser heat deposition to vaporization enthalpy contained in the vapor bubble.
 	}
 	\label{fig:efficiency}
 \end{figure*}

The conversion efficiency of light energy absorbed by the NP to vapour generation is crucial for various applications. The efficiency can be estimated as a ratio of the energy stored in the bubble and the deposited energy. Using the ideal gas law and Antoine's equation we obtain \cite{wang2018}:

\begin{equation}
	ef = \dfrac{E_{bubble}}{E_{deposited}} = \Lambda_{vap} \dfrac{M P_{sat} V_{max}}{R_g (\frac{B}{A-lg P_{sat}}-C) \xi P_l\tau_d} ,
\end{equation}

\noindent where $P_{sat}$ is saturation pressure, $M$ is the molar mass, $\Lambda_{vap}$ is the latent heat of vaporization, $R_g$ = 8.314 $J mol^{-1}K^{-1}$ is the universal gas constant, A, B and C are component-specific constants, and $\xi$ is the nanoparticle absorption coefficient.
The estimated efficiencies for all alkanes as functions of the deposited energy are shown in Figure \ref{fig:efficiency}. The efficiency decreases with increasing chain length, which means that for longer chains more energy is lost for bulk liquid heating. All curves in Fig. \ref{fig:efficiency} follow the same trend:
starting from a low deposited energy, which corresponds to short delay times and high laser powers, the efficiency goes up, reaches a maximum, and then rapidly decrease with increasing deposited energy, i.e. longer delay times and lower laser powers. The initial efficiency increase is caused by the heat diffusion dynamics (see Supporting Information). At the time scale of tens of $\mu s$ the heat diffusion is close to the 1D planar case, while later on the transition from planar heat diffusion to spherical heat diffusion occurs. Further increase of the deposited energy reflects long delays prior to nucleation, consequently, there is enough time for heat to diffuse deep into the bulk liquid. Therefore, the energy losses increase and the conversion efficiency quickly drops.

As already mentioned in the introduction, in water, there are four distinctly different phases of plasmonic microbubbles evolution \cite{wang2017,wang2018}. Up to now we only discussed the first one. For water, this first bubble phase is analogous to the here discussed first phase of alkanes and involves the formation of an explosive vapor bubble. \cite{wang2018}

However, in the case of water, the first phase is followed by a second oscillating bubble phase. In the case of alkanes this second phase is completely absent (see Movies 1 (a-c) in Supporting Information), which we ascribe to the fact that air solubility in alkanes is 15-30 times higher than in water (see Table \ref{properties}). Therefore, the bubble, which is generated by liquid evaporation right after the collapse of the first explosive bubble, already contains some gas and has more chances to survive and stabilize.

Evidence for the existence of gas in the bubble for the n-alkane case comes from the peculiar bubble behavior during the bubble collapse, which is especially pronounced for n-pentane (See Movie 1 (a) in Supporting Information). Instead of complete shrinkage, the bubble exhibits some sort of rebound, retains a certain size with a wavy interface and then detaches from the substrate, rising in the cool liquid medium. This dynamics is quite intriguing by itself but falls out of the scope of this paper. However, it is remarkable that the bubble does not vanish or condense immediately in this environment. Also the phenomenon of a stable rising bubble clearly suggests that the bubble formed after the collapse (or partial collapse in the case of n-pentane) of the initial explosive bubble is indeed partially filled with gas.

 \begin{figure*}[h]
 	\includegraphics[width=1\textwidth]{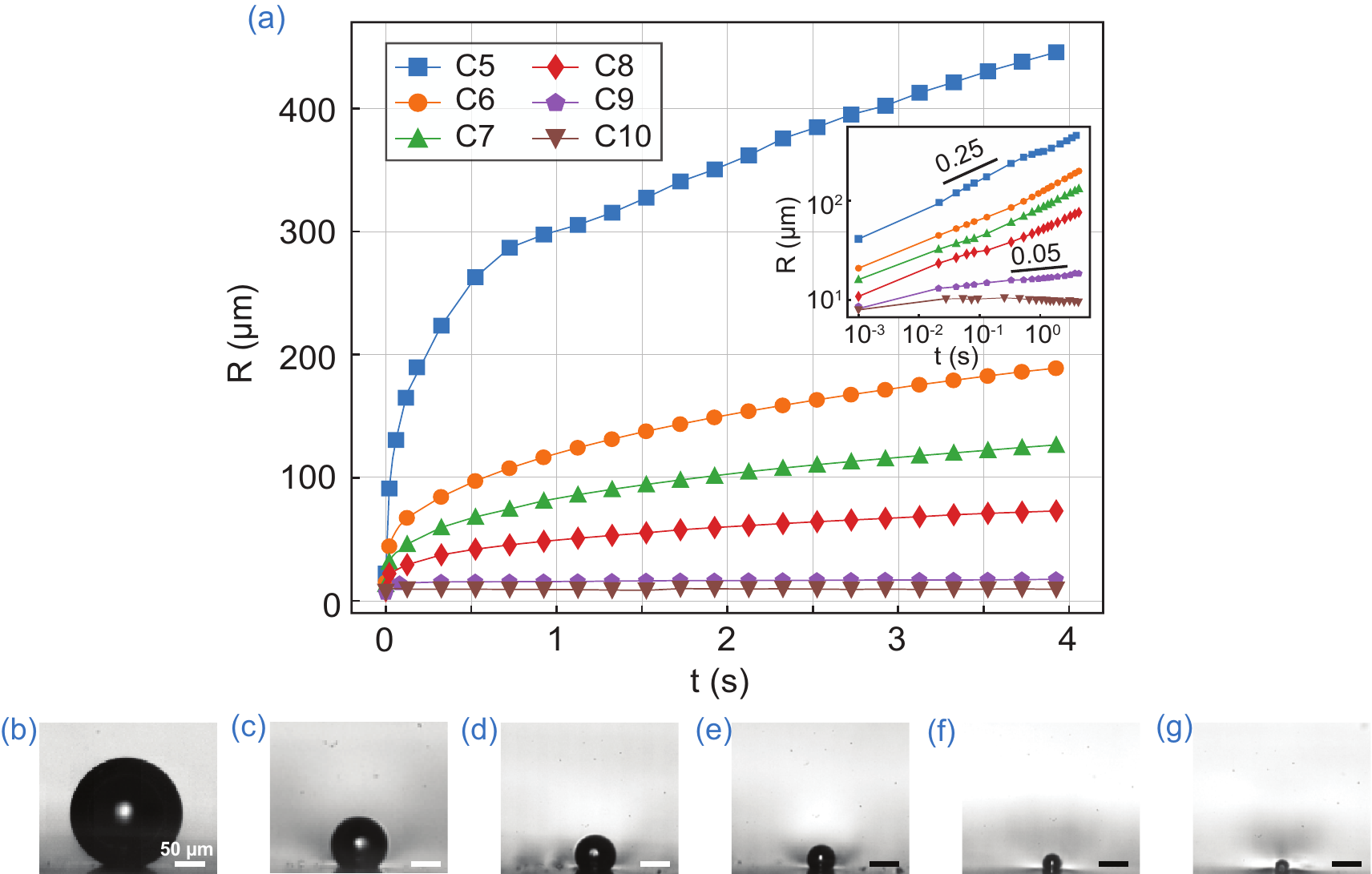}
 	\caption{
a) Radius versus time during plasmonic bubble growth in n-alkanes at 100 mW laser power. The inset shows the same data as double logarithmic plot with two slopes given as orientation; (b-g) Snapshots of the steadily growing bubble in n-alkanes: b) $C_5H_{12}$, c) $C_6H_{14}$, d) $C_7H_{16}$, e) $C_8H_{18}$, f) $C_9H_{20}$, g) $C_{10}H_{22}$, all after 20 ms of laser irradiation.
 	}
 	\label{fig:long term}
 \end{figure*}

During the second phase the bubble steadily grows. In Figure \ref{fig:long term} (a) the bubble radius versus time is shown for the first 4 seconds of laser irradiation at a laser power of 100 mW. The inset shows a double logarithmic plot of the bubble radius versus time. It is immediately clear from Figure \ref{fig:long term} that the longer the carbon chain of the alkane, the smaller the bubble. Whilst in $C_5H_{14}$, $C_6H_{14}$, $C_7H_{16}$ and $C_8H_{18}$, the bubble grows steadily with time, the growth of $C_9H_{20}$ and $C_{10}H_{22}$ bubbles quickly saturates.
From the data presented in Fig.\ref{fig:long term}, one can calculate the energy stored in the vapor bubble $E_{required} = \rho_vV_v(C_p \Delta T + \Lambda_{vap})$ for any alkane at any given time, for a laser power of 100 mW. The energy stored in the pentane bubble is an order of magnitude larger than that stored in the decane bubble. Remarkably, this result for the second phase of the bubble evolution mirrors the efficiency calculated in Fig. \ref{fig:efficiency} for the first phase. 
Power losses through the bubble interface might be estimated as $\mathcal{P} = 4 \pi R^2 \lambda \frac{T_{boil}-T_0}{\delta}$, where $\delta$ is thermal boundary layer thickness, and $T_0$ is room temperature (20$^\circ$C). Given the similar heat diffusion coefficients and thermal boundary layer thicknesses, for the same bubble size the decane bubble experiences a much larger energy loss than the pentane one $\frac{\mathcal{P}^{C10}}{\mathcal{P}^{C5}} \approx \frac{T_{boil}^{C10} - T_0}{T_{boil}^{C5} - T_0} \approx 10$.

\begin{figure*}[h]
 	\includegraphics[width=0.5\textwidth]{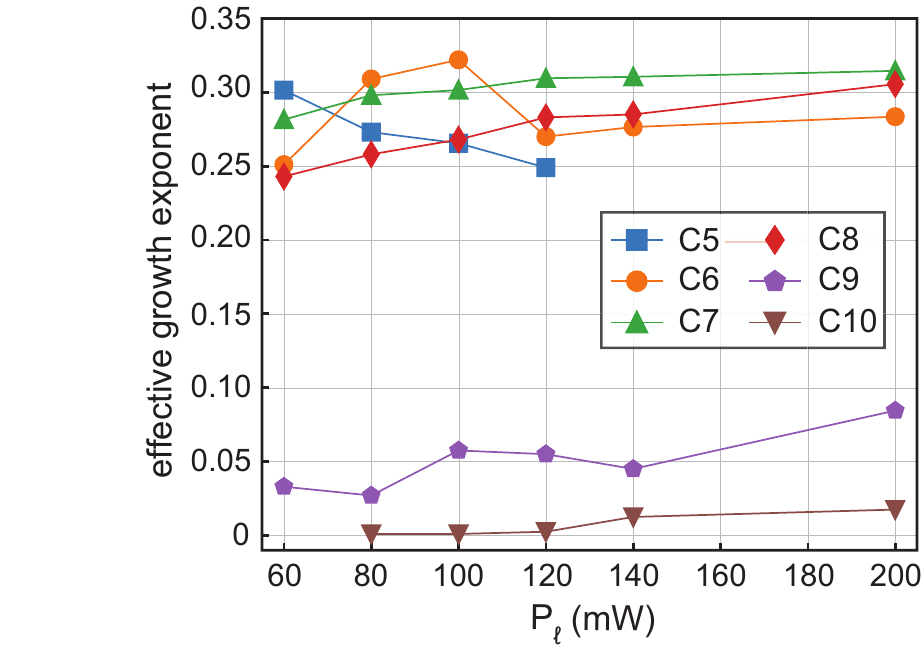}
 	\caption{
The effective growth exponent $\alpha$ of the bubble radius ($R \propto t^\alpha$) for pentane, hexane, heptane, octane, nonane and decane at different laser powers.
 	}
 	\label{fig:growth exponent}
 \end{figure*}

The effective growth exponent $\alpha$ of the bubble radius dynamics $R \propto t^\alpha$ for the n-alkanes versus laser power is  plotted in Figure \ref{fig:growth exponent}. Interestingly, the effective growth exponent of pentane, hexane, heptane, octane slightly varies from 0.25 to 0.30 for all laser powers, whereas for nonane and decane it is substantially smaller ($<$ 0.1). In the case of water the effective growth exponent is close to 0.33 and bubble growth is dominated by gas that is expelled from the liquid in the vicinity of the bubble, due to the decrease of gas solubility in water with increasing temperature \cite{wang2017}. In the case of n-alkanes, however, the solubility of gas increases with increasing temperature and therefore, the influx of dissolved gas is not the contributor to the second and final bubble life phase of alkanes. So the main source of the growth of the bubble is evaporation of the liquid and therefore, we conclude that the second life phase of plasmonic bubbles in n-alkanes is vaporization dominated.

\begin{figure*}[h]
 	\includegraphics[width=0.5\textwidth]{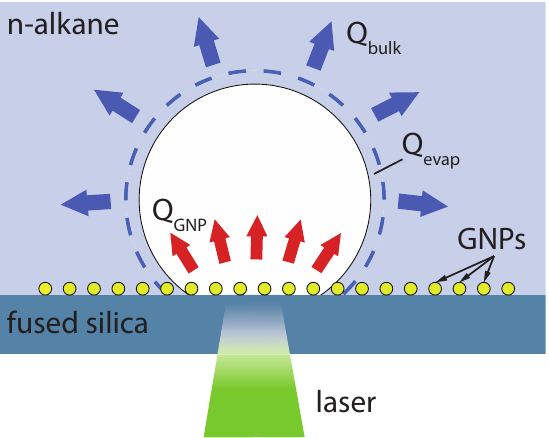}
 	\caption{
Schematic of the mechanism of bubble growth in n-alkanes on the long timescale.
 	}
 	\label{fig: heat interplay}
 \end{figure*}

In the second phase, the growth of the bubble is governed by a competition between heating from the nanoparticles and heat loss through the bubble interface as sketched in Figure \ref{fig: heat interplay}. Owing to the large diffusivity of the gas ($D_{dif} \approx 10^{-5}$ $m^2/s$) and to the slow dynamics at hands (seconds), the temperature in the gas is homogeneous. Additionally, the thermal boundary layer around the bubble is given sufficient time to fully develop ($\delta_{th} = R_b$). In first approximation, heat loss is therefore proportional to the bubble radius. As a result, the bubble will grow until the heat loss through the interface balances the thermal energy provided by the nanoparticles.
Solving two heat conduction equations corresponding to this steady state for the gaseous and liquid phase results in an estimated bubble size $R_{stable} = \frac{\xi P_i}{4 \pi \lambda_l (T_{boil} - T_0)} $, where $\xi$ is the nanoparticles absorption coefficient, $P_i$ is provided power, $\lambda_l$ is liquid conductivity and $T_0$ is room temperature (20 $^\circ$C).

For light alkanes with low boiling temperatures, the delivered thermal energy is sufficient to heat up and vaporize a large volume of liquid. This growth, however, is severely hindered for higher boiling temperatures as can be seen from Fig. \ref{fig:long term}(a) for $C_9H_{20}$ and $C_{10}H_{22}$.
It is worth to mention that despite a vaporization dominated growth during the second phase, in n-alkanes the bubbles are effectively composed of a gas/vapor mixture. Due to the high amount of dissolved gas in alkanes (15 to 30 times more than in water), a lot of gas is collected by the bubble during the vaporization process.

\section{Conclusion}
Plasmonic microbubbles in n-alkanes from $C_{5}H_{12}$ to $C_{10}H_{22}$ display two different phases in their evolution. The first phase occurs after a delay time $\tau_d$ after illumination of about 100 $\mu s$, when an explosive vapor microbubble is formed. The size of this bubble barely depends on the alkane chain length. The achievable local superheat of the alkane as well as light to vapor conversion efficiency decrease with elongating carbon chains. The second phase, which is also vaporization dominated, is characterized by a steadily growing bubble. The effective scaling law exponent $\alpha$ in $R(t)\propto t^\alpha$ is 0.25 - 0.30 for pentane, hexane, heptane and octane, whereas for nonane and decane it is substantially smaller ($<$0.1). Significant differences between plasmonic bubble dynamics in water and n-alkanes thus have been demonstrated, which is important for the design and regulation of the systems based on plasmonic bubble formation.

\begin{acknowledgement}

{The authors thank Bert Weckhuysen and Michel Versluis for helpful discussions and Dutch Organization for Research (NWO) and the Netherlands Center for Multiscale Catalytic Energy Conversion (MCEC) for financial support.}

\end{acknowledgement}

\begin{suppinfo}

Examples of bubbles nucleation and initial growth in pentane, heptane and nonane.

Maximum temperature evolution with time in three water-immersed idealized systems: a single sphere, an array of nanoparticles at the interface of 2 media, and an infinitely large heated plane for the same total deposited power.

\end{suppinfo}


\bibliography{biblio_alkanes}

\providecommand{\latin}[1]{#1}
\makeatletter
\providecommand{\doi}
  {\begingroup\let\do\@makeother\dospecials
  \catcode`\{=1 \catcode`\}=2 \doi@aux}
\providecommand{\doi@aux}[1]{\endgroup\texttt{#1}}
\makeatother
\providecommand*\mcitethebibliography{\thebibliography}
\csname @ifundefined\endcsname{endmcitethebibliography}
  {\let\endmcitethebibliography\endthebibliography}{}
\begin{mcitethebibliography}{35}
\providecommand*\natexlab[1]{#1}
\providecommand*\mciteSetBstSublistMode[1]{}
\providecommand*\mciteSetBstMaxWidthForm[2]{}
\providecommand*\mciteBstWouldAddEndPuncttrue
  {\def\EndOfBibitem{\unskip.}}
\providecommand*\mciteBstWouldAddEndPunctfalse
  {\let\EndOfBibitem\relax}
\providecommand*\mciteSetBstMidEndSepPunct[3]{}
\providecommand*\mciteSetBstSublistLabelBeginEnd[3]{}
\providecommand*\EndOfBibitem{}
\mciteSetBstSublistMode{f}
\mciteSetBstMaxWidthForm{subitem}{(\alph{mcitesubitemcount})}
\mciteSetBstSublistLabelBeginEnd
  {\mcitemaxwidthsubitemform\space}
  {\relax}
  {\relax}

\bibitem[Lapotko(2009)]{lapotko2009}
Lapotko,~D. Plasmonic Nanoparticle-Generated Photothermal Bubbles and Their
  Biomedical Applications. \emph{Nanomedicine} \textbf{2009}, \emph{4},
  813--845\relax
\mciteBstWouldAddEndPuncttrue
\mciteSetBstMidEndSepPunct{\mcitedefaultmidpunct}
{\mcitedefaultendpunct}{\mcitedefaultseppunct}\relax
\EndOfBibitem
\bibitem[Emelianov \latin{et~al.}(2009)Emelianov, Li, and
  O'Donnell]{emelianov2009}
Emelianov,~S.~Y.; Li,~P.-C.; O'Donnell,~M. Photoacoustics for Molecular Imaging
  and Therapy. \emph{Phys. Today} \textbf{2009}, \emph{62}, 34--39\relax
\mciteBstWouldAddEndPuncttrue
\mciteSetBstMidEndSepPunct{\mcitedefaultmidpunct}
{\mcitedefaultendpunct}{\mcitedefaultseppunct}\relax
\EndOfBibitem
\bibitem[Baffou and Quidant(2013)Baffou, and Quidant]{baffou2013}
Baffou,~G.; Quidant,~R. Thermo-Plasmonics: Using Metallic Nanostructures as
  Nano-Sources of Heat. \emph{Laser Photonics Rev.} \textbf{2013}, \emph{7},
  171--187\relax
\mciteBstWouldAddEndPuncttrue
\mciteSetBstMidEndSepPunct{\mcitedefaultmidpunct}
{\mcitedefaultendpunct}{\mcitedefaultseppunct}\relax
\EndOfBibitem
\bibitem[Shao \latin{et~al.}(2015)Shao, Xuan, Dai, Si, Li, and He]{shao2015}
Shao,~J.; Xuan,~M.; Dai,~L.; Si,~T.; Li,~J.; He,~Q. Near-Infrared-Activated
  Nanocalorifiers in Microcapsules: Vapor Bubble Generation for \textit {In
  Vivo} Enhanced Cancer Therapy. \emph{Angew. Chem., Int. Ed.} \textbf{2015},
  \emph{54}, 12782--12787\relax
\mciteBstWouldAddEndPuncttrue
\mciteSetBstMidEndSepPunct{\mcitedefaultmidpunct}
{\mcitedefaultendpunct}{\mcitedefaultseppunct}\relax
\EndOfBibitem
\bibitem[Liu \latin{et~al.}(2014)Liu, Fan, Ting, and Yeh]{liu2014}
Liu,~H.-L.; Fan,~C.-H.; Ting,~C.-Y.; Yeh,~C.-K. Combining Microbubbles and
  Ultrasound for Drug Delivery to Brain Tumors: Current Progress and Overview.
  \emph{Theranostics} \textbf{2014}, \emph{4}, 432--444\relax
\mciteBstWouldAddEndPuncttrue
\mciteSetBstMidEndSepPunct{\mcitedefaultmidpunct}
{\mcitedefaultendpunct}{\mcitedefaultseppunct}\relax
\EndOfBibitem
\bibitem[Fan \latin{et~al.}(2014)Fan, Hu, and Ohta]{fan2014}
Fan,~Q.; Hu,~W.; Ohta,~A.~T. Laser-Induced Microbubble Poration of Localized
  Single Cells. \emph{Lab Chip} \textbf{2014}, \emph{14}, 1572--1578\relax
\mciteBstWouldAddEndPuncttrue
\mciteSetBstMidEndSepPunct{\mcitedefaultmidpunct}
{\mcitedefaultendpunct}{\mcitedefaultseppunct}\relax
\EndOfBibitem
\bibitem[Neumann \latin{et~al.}(2013)Neumann, Urban, Day, Lal, Nordlander, and
  Halas]{neumann2013}
Neumann,~O.; Urban,~A.~S.; Day,~J.; Lal,~S.; Nordlander,~P.; Halas,~N.~J. Solar
  Vapor Generation Enabled by Nanoparticles. \emph{ACS Nano} \textbf{2013},
  \emph{7}, 42--49\relax
\mciteBstWouldAddEndPuncttrue
\mciteSetBstMidEndSepPunct{\mcitedefaultmidpunct}
{\mcitedefaultendpunct}{\mcitedefaultseppunct}\relax
\EndOfBibitem
\bibitem[Neumann \latin{et~al.}(2013)Neumann, Feronti, Neumann, Dong, Schell,
  Lu, Kim, Quinn, Thompson, Grady, Nordlander, Oden, and Halas]{neumann2013-1}
Neumann,~O.; Feronti,~C.; Neumann,~A.~D.; Dong,~A.; Schell,~K.; Lu,~B.;
  Kim,~E.; Quinn,~M.; Thompson,~S.; Grady,~N. \latin{et~al.}  Compact Solar
  Autoclave Based on Steam Generation Using Broadband Light-Harvesting
  Nanoparticles. \emph{Proc. Natl. Acad. Sci. U. S. A.} \textbf{2013},
  \emph{110}, 11677--11681\relax
\mciteBstWouldAddEndPuncttrue
\mciteSetBstMidEndSepPunct{\mcitedefaultmidpunct}
{\mcitedefaultendpunct}{\mcitedefaultseppunct}\relax
\EndOfBibitem
\bibitem[Fang \latin{et~al.}(2013)Fang, Zhen, Neumann, Polman, Javier Garcia~de
  Abajo, Nordlander, and Halas]{fang2013}
Fang,~Z.; Zhen,~Y.-R.; Neumann,~O.; Polman,~A.; Javier Garcia~de Abajo,~F.;
  Nordlander,~P.; Halas,~N.~J. Evolution of Light-Induced Vapor Generation at a
  Liquid-Immersed Metallic Nanoparticle. \emph{Nano Lett.} \textbf{2013},
  \emph{13}, 1736--1742\relax
\mciteBstWouldAddEndPuncttrue
\mciteSetBstMidEndSepPunct{\mcitedefaultmidpunct}
{\mcitedefaultendpunct}{\mcitedefaultseppunct}\relax
\EndOfBibitem
\bibitem[Boriskina \latin{et~al.}(2013)Boriskina, Ghasemi, and
  Chen]{boriskina2013}
Boriskina,~S.; Ghasemi,~H.; Chen,~G. Plasmonic Materials for Energy: From
  Physics to Applications. \emph{Mater. Today} \textbf{2013}, \emph{16},
  375--386\relax
\mciteBstWouldAddEndPuncttrue
\mciteSetBstMidEndSepPunct{\mcitedefaultmidpunct}
{\mcitedefaultendpunct}{\mcitedefaultseppunct}\relax
\EndOfBibitem
\bibitem[Ghasemi \latin{et~al.}(2014)Ghasemi, Ni, Marconnet, Loomis, Yerci,
  Miljkovic, and Chen]{ghasemi2014}
Ghasemi,~H.; Ni,~G.; Marconnet,~A.~M.; Loomis,~J.; Yerci,~S.; Miljkovic,~N.;
  Chen,~G. Solar Steam Generation by Heat Localization. \emph{Nat. Commun.}
  \textbf{2014}, \emph{5}\relax
\mciteBstWouldAddEndPuncttrue
\mciteSetBstMidEndSepPunct{\mcitedefaultmidpunct}
{\mcitedefaultendpunct}{\mcitedefaultseppunct}\relax
\EndOfBibitem
\bibitem[Baral \latin{et~al.}(2014)Baral, Green, Livshits, Govorov, and
  Richardson]{baral2014}
Baral,~S.; Green,~A.~J.; Livshits,~M.~Y.; Govorov,~A.~O.; Richardson,~H.~H.
  Comparison of Vapor Formation of Water at the Solid/Water Interface to
  Colloidal Solutions Using Optically Excited Gold Nanostructures. \emph{ACS
  Nano} \textbf{2014}, \emph{8}, 1439--1448\relax
\mciteBstWouldAddEndPuncttrue
\mciteSetBstMidEndSepPunct{\mcitedefaultmidpunct}
{\mcitedefaultendpunct}{\mcitedefaultseppunct}\relax
\EndOfBibitem
\bibitem[Guo \latin{et~al.}(2017)Guo, Fu, Wang, and Wang]{guo2017}
Guo,~A.; Fu,~Y.; Wang,~G.; Wang,~X. Diameter Effect of Gold Nanoparticles on
  Photothermal Conversion for Solar Steam Generation. \emph{RSC Adv.}
  \textbf{2017}, \emph{7}, 4815--4824\relax
\mciteBstWouldAddEndPuncttrue
\mciteSetBstMidEndSepPunct{\mcitedefaultmidpunct}
{\mcitedefaultendpunct}{\mcitedefaultseppunct}\relax
\EndOfBibitem
\bibitem[Krishnan \latin{et~al.}(2009)Krishnan, Park, and
  Erickson]{krishnan2009}
Krishnan,~M.; Park,~J.; Erickson,~D. Optothermorheological Flow Manipulation.
  \emph{Opt. Lett.} \textbf{2009}, \emph{34}, 1976--1978\relax
\mciteBstWouldAddEndPuncttrue
\mciteSetBstMidEndSepPunct{\mcitedefaultmidpunct}
{\mcitedefaultendpunct}{\mcitedefaultseppunct}\relax
\EndOfBibitem
\bibitem[Zhang \latin{et~al.}(2011)Zhang, Jian, Zhang, Wang, Li, and
  Tam]{zhang2011}
Zhang,~K.; Jian,~A.; Zhang,~X.; Wang,~Y.; Li,~Z.; Tam,~H.-y. Laser-Induced
  Thermal Bubbles for Microfluidic Applications. \emph{Lab Chip} \textbf{2011},
  \emph{11}, 1389--1395\relax
\mciteBstWouldAddEndPuncttrue
\mciteSetBstMidEndSepPunct{\mcitedefaultmidpunct}
{\mcitedefaultendpunct}{\mcitedefaultseppunct}\relax
\EndOfBibitem
\bibitem[Zhao \latin{et~al.}(2014)Zhao, Xie, Mao, Zhao, Rufo, Yang, Guo, Mai,
  and Huang]{zhao2014}
Zhao,~C.; Xie,~Y.; Mao,~Z.; Zhao,~Y.; Rufo,~J.; Yang,~S.; Guo,~F.; Mai,~J.~D.;
  Huang,~T.~J. Theory and Experiment on Particle Trapping and Manipulation via
  Optothermally Generated Bubbles. \emph{Lab Chip} \textbf{2014}, \emph{14},
  384--391\relax
\mciteBstWouldAddEndPuncttrue
\mciteSetBstMidEndSepPunct{\mcitedefaultmidpunct}
{\mcitedefaultendpunct}{\mcitedefaultseppunct}\relax
\EndOfBibitem
\bibitem[Tantussi \latin{et~al.}(2018)Tantussi, Messina, Capozza, Dipalo,
  Lovato, and De~Angelis]{tantussi2018}
Tantussi,~F.; Messina,~G.~C.; Capozza,~R.; Dipalo,~M.; Lovato,~L.;
  De~Angelis,~F. Long-Range Capture and Delivery of Water-Dispersed
  Nano-objects by Microbubbles Generated on 3D Plasmonic Surfaces. \emph{ACS
  Nano} \textbf{2018}, \emph{12}, 4116--4122\relax
\mciteBstWouldAddEndPuncttrue
\mciteSetBstMidEndSepPunct{\mcitedefaultmidpunct}
{\mcitedefaultendpunct}{\mcitedefaultseppunct}\relax
\EndOfBibitem
\bibitem[Xie and Zhao(2017)Xie, and Zhao]{xie2017}
Xie,~Y.; Zhao,~C. An Optothermally Generated Surface Bubble and its
  Applications. \emph{Nanoscale} \textbf{2017}, \emph{9}, 6622--6631\relax
\mciteBstWouldAddEndPuncttrue
\mciteSetBstMidEndSepPunct{\mcitedefaultmidpunct}
{\mcitedefaultendpunct}{\mcitedefaultseppunct}\relax
\EndOfBibitem
\bibitem[Baffou and Quidant(2014)Baffou, and Quidant]{baffou2014}
Baffou,~G.; Quidant,~R. Nanoplasmonics for Chemistry. \emph{Chem. Soc. Rev.}
  \textbf{2014}, \emph{43}, 3898--3907\relax
\mciteBstWouldAddEndPuncttrue
\mciteSetBstMidEndSepPunct{\mcitedefaultmidpunct}
{\mcitedefaultendpunct}{\mcitedefaultseppunct}\relax
\EndOfBibitem
\bibitem[Adleman \latin{et~al.}(2009)Adleman, Boyd, Goodwin, and
  Psaltis]{adleman2009}
Adleman,~J.~R.; Boyd,~D.~A.; Goodwin,~D.~G.; Psaltis,~D. Heterogenous Catalysis
  Mediated by Plasmon Heating. \emph{Nano Letters} \textbf{2009}, \emph{9},
  4417--4423\relax
\mciteBstWouldAddEndPuncttrue
\mciteSetBstMidEndSepPunct{\mcitedefaultmidpunct}
{\mcitedefaultendpunct}{\mcitedefaultseppunct}\relax
\EndOfBibitem
\bibitem[Lin \latin{et~al.}(2016)Lin, Peng, Mao, Li, Yogeesh, Rajeeva, Perillo,
  Dunn, Akinwande, and Zheng]{lin2016}
Lin,~L.; Peng,~X.; Mao,~Z.; Li,~W.; Yogeesh,~M.~N.; Rajeeva,~B.~B.;
  Perillo,~E.~P.; Dunn,~A.~K.; Akinwande,~D.; Zheng,~Y. Bubble-Pen Lithography.
  \emph{Nano Letters} \textbf{2016}, \emph{16}, 701--708\relax
\mciteBstWouldAddEndPuncttrue
\mciteSetBstMidEndSepPunct{\mcitedefaultmidpunct}
{\mcitedefaultendpunct}{\mcitedefaultseppunct}\relax
\EndOfBibitem
\bibitem[Baffou \latin{et~al.}(2014)Baffou, Polleux, Rigneault, and
  Monneret]{baffou2014jpc}
Baffou,~G.; Polleux,~J.; Rigneault,~H.; Monneret,~S. Super-Heating and
  Micro-Bubble Generation around Plasmonic Nanoparticles under cw Illumination.
  \emph{J. Phys. Chem. C} \textbf{2014}, \emph{118}, 4890--4898\relax
\mciteBstWouldAddEndPuncttrue
\mciteSetBstMidEndSepPunct{\mcitedefaultmidpunct}
{\mcitedefaultendpunct}{\mcitedefaultseppunct}\relax
\EndOfBibitem
\bibitem[Carlson \latin{et~al.}(2012)Carlson, Green, and
  Richardson]{carlson2012}
Carlson,~M.~T.; Green,~A.~J.; Richardson,~H.~H. Superheating Water by CW
  Excitation of Gold Nanodots. \emph{Nano Lett.} \textbf{2012}, \emph{12},
  1534--1537\relax
\mciteBstWouldAddEndPuncttrue
\mciteSetBstMidEndSepPunct{\mcitedefaultmidpunct}
{\mcitedefaultendpunct}{\mcitedefaultseppunct}\relax
\EndOfBibitem
\bibitem[Liu \latin{et~al.}(2015)Liu, Bao, Dipalo, De~Angelis, and
  Zhang]{liu2015}
Liu,~X.; Bao,~L.; Dipalo,~M.; De~Angelis,~F.; Zhang,~X. Formation and
  Dissolution of Microbubbles on Highly-Ordered Plasmonic Nanopillar Arrays.
  \emph{Sci. Rep.} \textbf{2015}, \emph{5}, 18515\relax
\mciteBstWouldAddEndPuncttrue
\mciteSetBstMidEndSepPunct{\mcitedefaultmidpunct}
{\mcitedefaultendpunct}{\mcitedefaultseppunct}\relax
\EndOfBibitem
\bibitem[Wang \latin{et~al.}(2017)Wang, Zaytsev, The, Eijkel, Zandvliet, Zhang,
  and Lohse]{wang2017}
Wang,~Y.; Zaytsev,~M.~E.; The,~H.~L.; Eijkel,~J. C.~T.; Zandvliet,~H. J.~W.;
  Zhang,~X.; Lohse,~D. Vapor and Gas-Bubble Growth Dynamics around
  Laser-Irradiated, Water-Immersed Plasmonic Nanoparticles. \emph{ACS Nano}
  \textbf{2017}, \emph{11}, 2045--2051\relax
\mciteBstWouldAddEndPuncttrue
\mciteSetBstMidEndSepPunct{\mcitedefaultmidpunct}
{\mcitedefaultendpunct}{\mcitedefaultseppunct}\relax
\EndOfBibitem
\bibitem[Wang \latin{et~al.}(2018)Wang, Zaytsev, Lajoinie, The, Eijkel, van~den
  Berg, Versluis, Weckhuysen, Zhang, Zandvliet, and Lohse]{wang2018}
Wang,~Y.; Zaytsev,~M.~E.; Lajoinie,~G.; The,~H.~L.; Eijkel,~J. C.~T.; van~den
  Berg,~A.; Versluis,~M.; Weckhuysen,~B.~M.; Zhang,~X.; Zandvliet,~H. J.~W.
  \latin{et~al.}  Giant and Explosive Plasmonic Bubbles by Delayed Nucleation.
  \emph{Proceedings of the National Academy of Sciences} \textbf{2018},
  \emph{115}, 7676--7681\relax
\mciteBstWouldAddEndPuncttrue
\mciteSetBstMidEndSepPunct{\mcitedefaultmidpunct}
{\mcitedefaultendpunct}{\mcitedefaultseppunct}\relax
\EndOfBibitem
\bibitem[Linstrom and Mallard(2018)Linstrom, and Mallard]{nist}
Linstrom,~P.~J.; Mallard,~W.~G. \emph{NIST Chemistry WebBook, NIST Standard
  Reference Database Number 69}; 2018\relax
\mciteBstWouldAddEndPuncttrue
\mciteSetBstMidEndSepPunct{\mcitedefaultmidpunct}
{\mcitedefaultendpunct}{\mcitedefaultseppunct}\relax
\EndOfBibitem
\bibitem[Le-The \latin{et~al.}(2017)Le-The, Berenschot, Tiggelaar, Tas, van~den
  Berg, and Eijkel]{the2017}
Le-The,~H.; Berenschot,~E.; Tiggelaar,~R.~M.; Tas,~N.~R.; van~den Berg,~A.;
  Eijkel,~J. C.~T. Shrinkage Control of Photoresist for Large-Area Fabrication
  of Sub-30 nm Periodic Nanocolumns. \emph{Advanced Materials Technologies}
  \textbf{2017}, \emph{2}, 1600238\relax
\mciteBstWouldAddEndPuncttrue
\mciteSetBstMidEndSepPunct{\mcitedefaultmidpunct}
{\mcitedefaultendpunct}{\mcitedefaultseppunct}\relax
\EndOfBibitem
\bibitem[Skripov(1974)]{skripov1974}
Skripov,~V.~P. \emph{Metastable Liquids}; Wiley, 1974\relax
\mciteBstWouldAddEndPuncttrue
\mciteSetBstMidEndSepPunct{\mcitedefaultmidpunct}
{\mcitedefaultendpunct}{\mcitedefaultseppunct}\relax
\EndOfBibitem
\bibitem[Skripov and Pavlov(1970)Skripov, and Pavlov]{skripov1970}
Skripov,~V.~P.; Pavlov,~P.~A. Explosive Boiling of Liquids and Fluctuation
  Nucleus Formation. \emph{High Temp.} \textbf{1970}, \emph{8}, 782\relax
\mciteBstWouldAddEndPuncttrue
\mciteSetBstMidEndSepPunct{\mcitedefaultmidpunct}
{\mcitedefaultendpunct}{\mcitedefaultseppunct}\relax
\EndOfBibitem
\bibitem[Blander and Katz(1975)Blander, and Katz]{blander1975}
Blander,~M.; Katz,~J.~L. Bubble Nucleation in Liquids. \emph{AIChE Journal}
  \textbf{1975}, \emph{21}, 833--848\relax
\mciteBstWouldAddEndPuncttrue
\mciteSetBstMidEndSepPunct{\mcitedefaultmidpunct}
{\mcitedefaultendpunct}{\mcitedefaultseppunct}\relax
\EndOfBibitem
\bibitem[Puchinskis and Skripov(2001)Puchinskis, and Skripov]{puchinskis2001}
Puchinskis,~S.~E.; Skripov,~P.~V. The Attainable Superheat: From Simple to
  Polymeric Liquids. \emph{International Journal of Thermophysics}
  \textbf{2001}, \emph{22}, 1755--1768\relax
\mciteBstWouldAddEndPuncttrue
\mciteSetBstMidEndSepPunct{\mcitedefaultmidpunct}
{\mcitedefaultendpunct}{\mcitedefaultseppunct}\relax
\EndOfBibitem
\bibitem[Battino \latin{et~al.}(1984)Battino, Rettich, and
  Tominaga]{battino1984}
Battino,~R.; Rettich,~T.~R.; Tominaga,~T. The Solubility of Nitrogen and Air in
  Liquids. \emph{Journal of Physical and Chemical Reference Data}
  \textbf{1984}, \emph{13}, 563--600\relax
\mciteBstWouldAddEndPuncttrue
\mciteSetBstMidEndSepPunct{\mcitedefaultmidpunct}
{\mcitedefaultendpunct}{\mcitedefaultseppunct}\relax
\EndOfBibitem
\bibitem[Mizerovsky and Smirnova(2010)Mizerovsky, and Smirnova]{mizerovsky2010}
Mizerovsky,~L.~N.; Smirnova,~K.~P. Temperature Dependence of the Solubility of
  Nitrogen in n-Alkanes at Atmospheric Pressure. \emph{Russian Chemical
  Bulletin} \textbf{2010}, \emph{59}, 673--676\relax
\mciteBstWouldAddEndPuncttrue
\mciteSetBstMidEndSepPunct{\mcitedefaultmidpunct}
{\mcitedefaultendpunct}{\mcitedefaultseppunct}\relax
\EndOfBibitem
\end{mcitethebibliography}

\end{document}


\subsection*{Heat diffusion dynamics}

\begin{figure*}[h]
 	\includegraphics[width=0.5\textwidth]{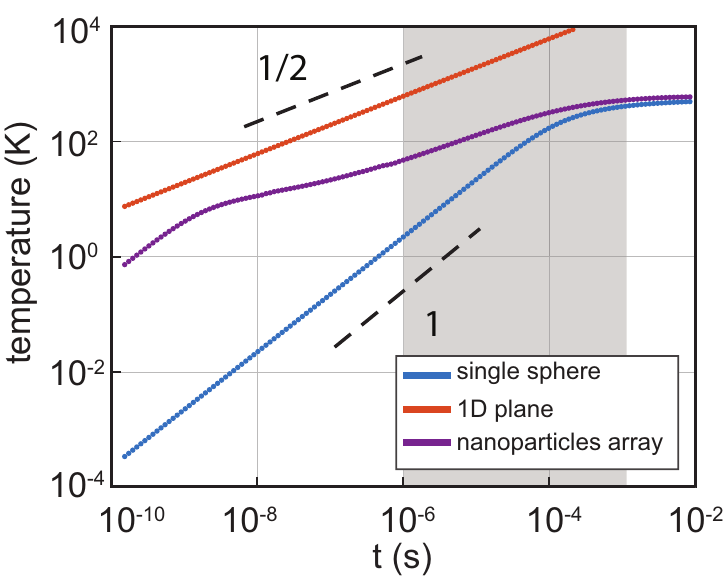}
 	\caption{
   Maximum temperature evolution with time in three water-immersed idealized systems: a single sphere (blue), an array of nanoparticles at the interface of 2 media (purple), and an infinitely large heated plane (orange) for the same total deposited power. The gray area represents the range of times in which the onset of vaporization was observed.
 	}
 	\label{fig:heat dynamics}
 \end{figure*}

The temperature evolution resulting from the same type of calculations as have been performed in our recent paper\textsuperscript{1} is shown in Figure \ref{fig:heat dynamics}. On small length scale and sub-microsecond timescale, the heat first diffuses spherically from each nanoparticle. Later on, the temperature fields from individual nanoparticles start to superpose, leading to a transition towards one 1D planar diffusing heat front. Finally, due to the finite size of the laser spot, heat diffusion becomes spherical again after a typical time $\tau = R_\ell^2/(\pi k)$, where $R_\ell$ the radius of the Gaussian laser beam. There, the temperature converges toward a steady state value. If the steady state temperature is below the nucleation temperature of liquid, vaporization cannot occur.
